\documentclass[doublecol]{epl2} 
\usepackage{amsmath, amssymb}
\usepackage{lipsum}
\usepackage{algorithm}
\usepackage{algpseudocode}
\usepackage{mathrsfs}
\usepackage{xcolor}
\usepackage{booktabs}
\usepackage{hyperref}
\usepackage[normalem]{ulem}
\usepackage{bm}

\hypersetup{
    colorlinks=true,
    linkcolor=blue,       
    filecolor=magenta,  
    urlcolor=purple,       
    citecolor=purple,         
    pdftitle={Stochastic closure},
    pdfpagemode=FullScreen,
}

\title{On the importance of stochasticity in closures of turbulence}
\shorttitle{On the importance of stochasticity in closures of turbulence} %Insert here a short version of the title if it exceeds 70 characters

\author{
A. Freitas\inst{1,2} \and
L. Biferale\inst{1} \and
M. Desbrun\inst{3} \and
G. Eyink\inst{4} \and
A. A. Mailybaev\inst{5} \and
K. Um\inst{2}
}
\shortauthor{A. Freitas \etal}

\institute{                    
  \inst{1} Dept. Physics and INFN, University of Rome “Tor Vergata”, Italy \\
  \inst{2} LTCI, Télécom Paris, IP Paris, France \\
  \inst{3} Inria and École Polytechnique, IP Paris, France \\
  \inst{4} Dept. of Applied Mathematics \& Statistics, The Johns Hopkins University, Baltimore, MD, USA \\
  \inst{5} Instituto de Matemática Pura e Aplicada – IMPA, Rio de Janeiro, Brazil \\
}
\pacs{nn.mm.xx}{First pacs description}
\pacs{nn.mm.xx}{Second pacs description}
\pacs{nn.mm.xx}{Third pacs description}

\abstract{
Deterministic closures for coarse-grained turbulence models help reproduce mean statistics, but often fail to capture the finite-time growth of uncertainty. Using the framework of shell models as a quantitative multi-scale testbed, we compare fully resolved simulations with large-eddy simulations using either stochastic or deterministic subgrid closures. While in the fully resolved system a single microscopic perturbation is rapidly amplified by strongly chaotic dynamics, truncation produces a strong delay and suppression of variance growth when uncertainty is introduced through initial condition perturbations only. We show that a data-driven Langevin-type stochastic closure restores the correct timing and magnitude of variance growth across scales, demonstrating that sustained stochasticity is essential for predictability in reduced turbulent dynamics.
}

\begin{document}

\maketitle

%-----------------------------------------
%-----------------------------------------

\section{Introduction}
\label{sec:intro}

Turbulence is a state of a physical system far from equilibrium, characterized by a large number of strongly interacting degrees of freedom spanning a wide range of spatial and temporal scales~\cite{frisch, alexakis2018, sreeni25}. In practice, this complexity makes fully resolved simulations infeasible at the high Reynolds numbers ($\mathrm{Re} = UL/\nu$, where $U$ and $L$ are a characteristic velocity and length, respectively, and $\nu$ is kinematic viscosity) of interest, and turbulent flows are therefore commonly studied using reduced-order models such as Large-Eddy Simulation (LES)~\cite{les_review}, where only the large scales are explicitly resolved. The effect of unresolved small-scale degrees of freedom on the resolved dynamics must then be modeled, giving rise to what is known as the closure problem. A schematic illustration of the problem is shown in Fig~\ref{fig:fig1}. Despite its central importance, this problem remains fundamentally challenging, as intermittency~\cite{frisch_1985, benzi_multifractal_84} and the lack of scale separation prevent a simple modelling in terms of slow (fully resolved) and fast (modelled) variables.

In many applications, such as weather and climate forecasting~\cite{hasselmann_stochastic_1976, palmer_stochastic_2019}, astrophysics and cosmology~\cite{keller_chaos_2019, genel_butterfly_2019}, the goal of turbulence modelling is not limited to the prediction of asymptotic steady states, but also predictability over finite, physically-relevant time horizons, starting from a given initial condition (or an ensemble thereof). Addressing this question requires capturing not only unconditional flow statistics, but also the evolution of uncertainty. Most LES closures, however, are deterministic by construction and are primarily designed to reproduce mean fluxes or spectra~\cite{lesieur_1996, les_review, Pope_2004}. In such deterministic LES, ensemble spread arises only through sensitivity to initial conditions. Here we show that this can severely misrepresent finite-time uncertainty growth, and that an explicit and suitable stochastic subgrid component can be essential to obtain faithful uncertainty evolution in coarse-grained predictions.

From a formal standpoint, the need to incorporate stochastic elements in reduced-order models is already apparent in the Mori--Zwanzig projection formalism~\cite{mori,zwanzig,MZ_2010}. When a high-dimensional nonlinear system is projected onto a subset of resolved variables, the exact reduced dynamics consists of a Markovian contribution, a memory term, and an effective stochastic forcing that accounts for unresolved--unresolved interactions. While explicit forms of these terms are generally intractable, this framework makes clear that stochastic forcing is not an ad-hoc modelling choice, but a structural component of any rigorous attempt to model the coarse-grained evolution. An equivalent conclusion follows from a path-integral formulation of the Navier--Stokes equations, where resolved and unresolved modes are separated by a spatial filter and the latter are integrated out exactly~\cite{eyink1996turbulence}. The resulting reduced dynamics takes the form of a generalized Langevin equation with memory and non-Gaussian noise, and already suggested modelling unresolved modes by self-consistent stochastic Langevin surrogates, providing an early conceptual basis for stochastic LES closures.

Predictability in chaotic systems was classically framed by Lorenz~\cite{lorenz63} in terms of exponential growth of small but finite initial perturbations. In a later work, however, Lorenz emphasized that in multiscale systems small-scale disturbances may propagate upscale in finite time through an \emph{inverse cascade of errors}~\cite{lorenz69}. In this setting, stochasticity at the level of the primitive equations can arise from a quasi-singular role played by (thermal) fluctuations at very small scales~\cite{lorenz69,Betchov_1957,betchov_1960,betchov_1964,leith_kraichnan_1972,ruelle79}, where even vanishingly weak perturbations may be rapidly amplified. This mechanism has been confirmed by direct numerical simulations based on fluctuating hydrodynamics and molecular gas dynamics, which show that thermal fluctuations influence the dissipation range and are obviously not captured by the deterministic Navier–Stokes equations~\cite{mcmullen_2022,Bell_Nonaka_Garcia_Eyink_2022}. More recently, shell model studies~\cite{bandak_2022,bandak_prl_2024} have demonstrated that such microscopic fluctuations can be transported to large scales within a single large-eddy turnover time.

These findings are closely connected to the phenomenon of spontaneous stochasticity~\cite{eyink_goldefeld_2025}, which refers to the emergence of intrinsically stochastic solutions with non-trivial (non-delta) probability distributions in the double limit $\mathrm{Re}^{-1}\to0$ and vanishing noise amplitude, reflecting a loss of uniqueness in the effective inviscid dynamics. It manifests in Lagrangian form as non-unique particle trajectories in rough velocity fields~\cite{richardson1926, bernard_slow_1998, Vanden_Eijnden_2000, falkovich_particles_2001, Kupiainen_2004}, and in Eulerian form as the emergence of macroscopically distinct field realizations~\cite{lorenz69, Mailybaev_2016,Mailybaev_2017, biferale_2018, thalabard_butterfly_2020, Drivas_2021,mailybaev_spontaneously_2023, DRIVAS_MAILYBAEV_RAIBEKAS_2024, dubrulle25}. Although formally defined as an asymptotic phenomenon, its implications extend to large but finite Reynolds numbers and finite noise amplitudes, where weak physical noise alone suffices to generate finite uncertainty on dynamically relevant time scales~\cite{bandak_prl_2024}. Recent analyses have emphasized that this mechanism may be suppressed in popular artificial-intelligence-based weather prediction models, which fail to reproduce the proverbial “butterfly effect” associated with spontaneous stochasticity~\cite{selz_craig_2023}. These observations further highlight the importance of explicitly representing sustained subgrid stochasticity in reduced models, regardless of whether the modelling approach is phenomenological or data-driven.

In this work, we investigate the importance of stochastic closures in turbulence using a controlled shell model~\cite{biferale2003shell} setting. We introduce a data-driven stochastic closure of Langevin type, in which the effect of unresolved scales is represented by machine-learned drift terms together with an explicit stochastic forcing. We assess its performance by comparison with deterministic closures with initial condition perturbations and with the equivalent in the shell model set up of a fully resolved  Landau--Lifshitz fluctuating hydrodynamics~\cite{landau_lifshitz_fluid_1959}. Focusing on ensemble variance growth of velocity at finite times, we show that deterministic closures with initial condition perturbations artificially suppress uncertainty, while our stochastic closures recover the correct temporal scaling up to the large-eddy turnover time.

%-----------------------------------------
%-----------------------------------------
\section{Fluctuating hydrodynamics}
\label{sec:fluct_hydro}
%-----------------------------------------
As a stochastic reference for the evolution of uncertainty, we consider a fluctuating hydrodynamics extension of the Sabra shell model~\cite{Lvov98} in the spirit of Landau--Lifshitz~\cite{landau_lifshitz_fluid_1959}. The aim is to use a simple perturbation --- acting far below the LES cutoff --- as a controlled benchmark for predictability studies. We consider complex shell velocities $u_n(t)\in\mathbb C$, $n=0,\dots,N-1$, associated with dyadic wavenumbers $k_n \!=\! 2^n$. The model is governed by a system of stochastic ordinary differential equations~\cite{bandak_2022}:
\begin{align}
\frac{\mathrm d u_n}{\mathrm dt}
&=
i\Bigl[
      k_{n+1} u_{n+2} u_{n+1}^\ast
      -\tfrac12 k_n u_{n+1} u_{n-1}^\ast
      \nonumber\\
&\
      +\tfrac12 k_{n-1} u_{n-1} u_{n-2}
    \Bigr]
- \nu k_n^2 u_n
+ f_n
+ \sqrt{\Theta}\,k_n\,\xi_n(t),
\label{eq:sabra_ll}
\end{align}
with coefficients $(a,b,c)=(1,-1/2,-1/2)$, conserving energy and a helicity-like invariant in the inviscid limit. The forcing $f_n$ is deterministic and acts only on the first two shells, $n\!=\!0,1$. We impose boundary conditions on the shells through  $u_{-2}\!=\!u_{-1}\!=\!u_{N+1}\!=\!u_{N+2}\!=\!0$.

The stochastic terms $\xi_n(t)$ are independent complex Gaussian white noises with covariance
\begin{equation}
\big\langle \xi_n^\ast(t)\,\xi_m(t') \big\rangle
=
2\,k_n^{\alpha}\,\delta_{nm}\,\delta(t-t'),
\qquad \alpha\in\{0,3\}.
\label{eq:noise_cov}
\end{equation}

The exponent $\alpha$ prescribes the spectral weighting of the microscopic forcing: $\alpha\!=\!0$ corresponds to shell-wise white noise with equal variance across $n$, while $\alpha\!=\!3$ increases the noise intensity towards small scales (closer, dimensionally, to the $3$D Landau--Lifshitz choice). Differences between $\alpha\!=\!0$ and $\alpha\!=\!3$ are confined to the thermalized far-dissipation tail well beyond the cutoff. The results reported here use $\alpha=0$; we have confirmed that using $\alpha=3$ does not affect the results or conclusions.

The nondimensional noise amplitude is parameterised as~\cite{bandak_2022, bandak_prl_2024}
\begin{equation}
\Theta
=
\mathrm{Re}^{-\beta}\,\theta_\eta,
\qquad
\beta=\frac{3}{4}(\alpha+2),
\qquad
\mathrm{Re}=\nu^{-1},
\label{eq:theta}
\end{equation}
so that the stochastic forcing vanishes as $\mathrm{Re}\!\to\!\infty$. Here
\begin{equation}
\theta_\eta=\frac{k_B T}{\rho\,u_\eta^2\,\eta^3},
\end{equation}
measures the ratio of thermal energy fluctuations to kinetic energy fluctuations at the Kolmogorov scale, where $k_B$ is the Boltzmann constant, $T$ temperature and $\rho$ density. Its precise magnitude is not crucial for our conclusions: it primarily controls the amplitude of fluctuations in the far-dissipation range, while the main quantity of interest here, the inertial range growth of uncertainty, is rapidly dominated by nonlinear amplification, becoming largely insensitive to the microscopic noise level. The parameter values used throughout are specified in the caption of Fig.~\ref{fig:fig1}.

Details of the numerical integration scheme are provided in Appendix A. 

%-----------------------------------------
%-----------------------------------------
%-----------------------------------------
\section{Closure problem}
\label{sec:closure}

Large-eddy simulation (LES) resolves only the large scales and therefore requires a closure to represent the effect of eliminated degrees of freedom. In the present setting, the Landau--Lifshitz reference injects microscopic fluctuations deep in the dissipation range; although these fluctuations are negligible in terms of mean energy at resolved scales, they can be rapidly amplified and transported upscale, directly impacting predictability. A reduced model that is deterministic by construction cannot represent this continual injection of subgrid uncertainty, motivating an explicit stochastic component in the closure. More fundamentally, even in a purely deterministic system, unresolved scales should be modelled stochastically, consistent with the Mori–Zwanzig formalism outlined in the Introduction.

\begin{figure}[htb]
    \centering
    \includegraphics[width=0.99\linewidth]{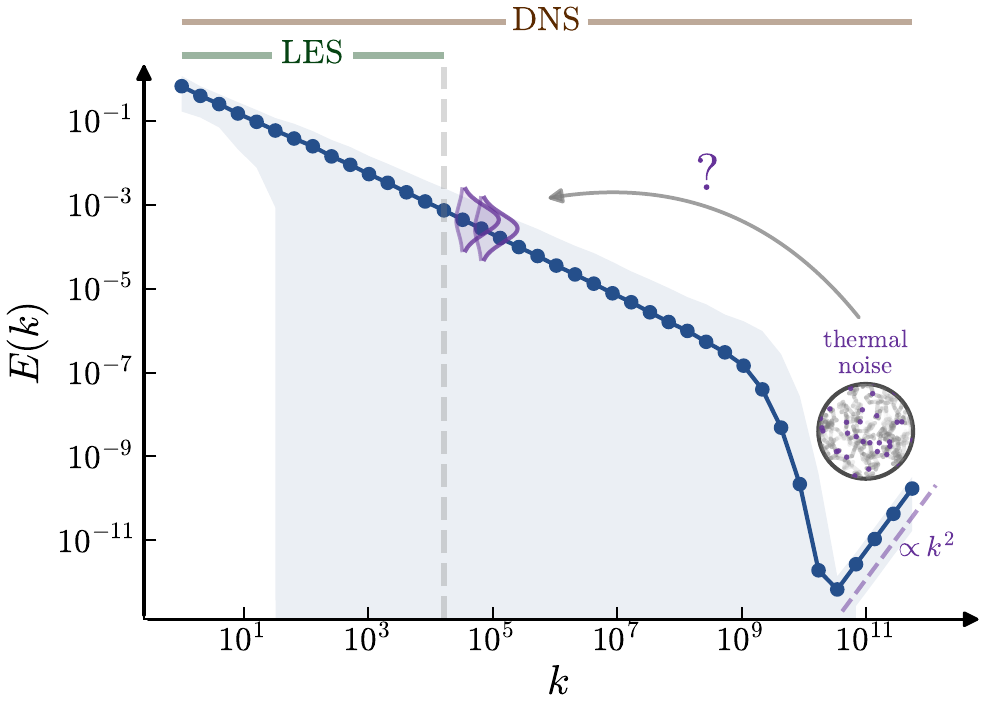} \vspace*{-8mm}
    \caption{\textbf{Problem setup}. 
    Energy spectrum $E(k)$ as a function of wavenumber $k$ obtained from Landau--Lifshitz fluctuating hydrodynamics (DNS). Large-eddy simulations (LES) resolve the dynamics only up to a cutoff scale $k_c$, while smaller scales are unresolved and must be modeled through a closure. Although thermal fluctuations dominate only in the far-dissipation range, their influence can propagate upscale and affect predictability at resolved scales. In shell models, where nonlinear interactions are local in scale, the closure problem reduces to modelling the two shells after the cutoff, including their stochastic statistics. Throughout this work, we use $N\!=\!40$, $\mathrm{Re}\!=\!10^{12}$, and $\theta_\eta \!=\! 2.83\times10^{-8}$, yielding a well-developed inertial range and a clean separation between near and far dissipation ranges. The value of $\theta_\eta$ used is the same as in Refs.~\cite{bandak_2022,bandak_prl_2024}}
    \label{fig:fig1}
\end{figure}

When the Sabra shell model is truncated at shell $s$, the last two resolved equations ($n\!=\!s-1,s$) depend on the two unresolved shells $u_{s+1}$ and $u_{s+2}$~\cite{freitas24,freitas25,JDL24,JDL25}. We close the reduced system by modelling these shells $(\hat u_{s+1},\hat u_{s+2})$ with a neural network~\cite{Goodfellow-et-al-2016} and inserting them into the missing triads. The problem setup is depicted in Fig.~\ref{fig:fig1}. The fluctuating hydrodynamics model (Eq.~\eqref{eq:sabra_ll}) introduced above is not used for training the neural network LES subgrid closure. Instead, training is carried out on deterministic Sabra data since the closure is learned from resolved scale dynamics alone, without direct access to the imposed microscopic forcing. The Landau--Lifshitz model is used exclusively as a benchmark to test whether a closure can reproduce the correct finite-time propagation of uncertainty when microscopic perturbations are injected far below the cutoff.

Let $u^{<}(t)\!=\!(u_0,\dots,u_s)(t)$ denote the resolved state. The network input is the last three resolved shells (fixing the flux locally),
\begin{equation}
x(t)=\bigl(u_{s-2}(t),u_{s-1}(t),u_s(t)\bigr)\in\mathbb C^3,
\label{eq:nn_input}
\end{equation}
and the network outputs a drift term for the two unresolved shells,
\begin{equation}
b_\theta\!\bigl(x(t)\bigr)=\bigl(b_{\theta,s+1}(x(t)),\,b_{\theta,s+2}(x(t))\bigr)\in\mathbb C^2,
\label{eq:nn_drift}
\end{equation}
where $\theta$ denotes that this is a neural network output. We use a multi-layer perceptron~\cite{Goodfellow-et-al-2016} with ReLU activations (7 hidden layers, width 256) and a linear output layer.

We model $(\hat u_{s+1},\hat u_{s+2})$ as a Markovian Langevin process driven by the learned drift and an explicit stochastic forcing,
\begin{equation}
\mathrm d\hat u_{n}(t)
=
b_{\theta,n}\!\bigl(x(t)\bigr)\,\mathrm dt
+
\sigma_{n}\,\mathrm dW_{n}(t),
\qquad n=s+1,s+2,
\label{eq:subgrid_langevin}
\end{equation}
where $W_{s+1},W_{s+2}$ are independent complex Wiener processes with normalisation
$\langle \mathrm dW_n^\ast\,\mathrm dW_m\rangle =2\,\delta_{nm}\,\mathrm dt$. We denote by $\Delta t$ the fine time step used for the Landau--Lifshitz/DNS reference, and by $\Delta\tilde t$ the coarse time step of the reduced LES model (with $\Delta\tilde t \!\gg\! \Delta t$; here $\Delta\tilde t\!=\!10^{3}\Delta t$). At time step $\Delta\tilde t$, we implement Eq.~\eqref{eq:subgrid_langevin} by Euler- Maruyama~\cite{kloeden1992nsde},
\begin{equation}
\hat u_n(t+\Delta\tilde t)
=
\hat u_n(t)
+
\Delta\tilde t\,b_{\theta,n}\!\bigl(x(t)\bigr)
+
\sqrt{\Delta\tilde t}\,\sigma_n\,\xi_{n,t},
\label{eq:subgrid_em}
\end{equation}
with independent complex Gaussians $\xi_{n,t}\sim\mathcal N_{\mathbb C}(0,1)$ so that $\langle|\xi_{n,t}|^2\rangle=2$.

For simplicity, the diffusion amplitudes $\sigma_n$ are not learned, but modeled based on dimensional scaling through
\begin{equation}
\sigma_n
=
\alpha_\sigma\,k_n^{1/2}
\left\langle |u_n^{\mathrm{DNS}}|^2 \right\rangle^{3/4},
\label{eq:sigma_subgrid}
\end{equation}
where $\alpha_\sigma$ is a dimensionless constant and $\langle\cdot\rangle$ denotes a time/ensemble average in the fully resolved deterministic Sabra simulations. 

Given $(u^{<},\hat u_{s+1},\hat u_{s+2})$, the closure contribution is assembled by inserting $\hat u_{s+1},\hat u_{s+2}$ into the missing triads in the Sabra nonlinearity, and the resolved state is advanced with one step of the numerical solver. Because training is performed by unrolling the reduced dynamics, the stochastic forcing in Eq.~\eqref{eq:subgrid_em} influences the optimization implicitly through its effect on multi-step trajectories.

We train $b_\theta$ \emph{a posteriori}~\cite{um2020sol, freitas24, freitas25} by unrolling the reduced stochastic model for $m\Delta t$ steps and matching the resolved trajectory to the deterministic reference at the same times. Denote by $u^{<}_{\theta}(t_m)$ the resolved trajectory generated by the reduced model and by $u^{<}_{\mathrm{ref}}(t_m)$ the resolved projection of the deterministic reference, at times $t_m=t_0+m\Delta\tilde t$. Our normalised multi-step loss is
\begin{equation}
\mathcal L=\frac{1}{M\,N_{\mathrm{Loss}}}\sum_{m=1}^{M}\sum_{n=1}^{N_{\mathrm{Loss}}}
\frac{|u^{<}_{\theta,n}(t_m)-u^{<}_{\mathrm{ref},n}(t_m)|^2}
{\sqrt{|u^{<}_{\theta,n}(t_m)|^2} \sqrt{|u^{<}_{\mathrm{ref},n}(t_m)|^2}}\,,
\label{eq:loss}
\end{equation}
where $N_{\mathrm{Loss}}$ is the number of resolved shells included in the loss (here we consider six shells up to the cutoff). Parameters are optimized with Adam (learning rate $10^{-4}$); further details are given in Appendix~B.

%-----------------------------------------
%-----------------------------------------
\section{Results}
\label{sec:results}
%-----------------------------------------
We analyse how uncertainty grows and propagates by comparing fully resolved simulations with large-eddy simulations, with emphasis on the role of stochasticity in the closure. Uncertainty can be quantified by the ensemble variance of the shell velocities,
\begin{equation}
\mathrm{Var}[u_n(t)]
=
\big\langle \big|u_n(t)-\langle u_n(t)\rangle\big|^2 \big\rangle,
\label{eq:var_def}
\end{equation}
where $\langle\cdot\rangle$ denotes an average over an ensemble of independent realisations evolving from the same initial condition. For each initial condition, we generate an ensemble of $1024$ members (independent noise realisations in DNS (Eq~\eqref{eq:sabra_ll}), and independent closure noise in stochastic LES). Reported variance curves are further averaged over $256$ initial conditions, obtained by sampling snapshots from the statistically stationary regime of fully-resolved deterministic Sabra simulations and truncated to the resolved range. Note that an analogous study for the 3D Navier--Stokes equations at comparable Reynolds numbers is currently out of reach: beyond the cost of a single high-$\mathrm{Re}$ DNS, reliable variance estimates require very large ensembles. Here we perform $256\times1024\simeq2.6\times10^{5}$ realisations (each integrated over $\mathcal O(\tau_0)$), a computational load that would be prohibitive for fully resolved 3D turbulence.

\begin{figure*}[htb]
    \centering
    \includegraphics[width=0.99\linewidth]{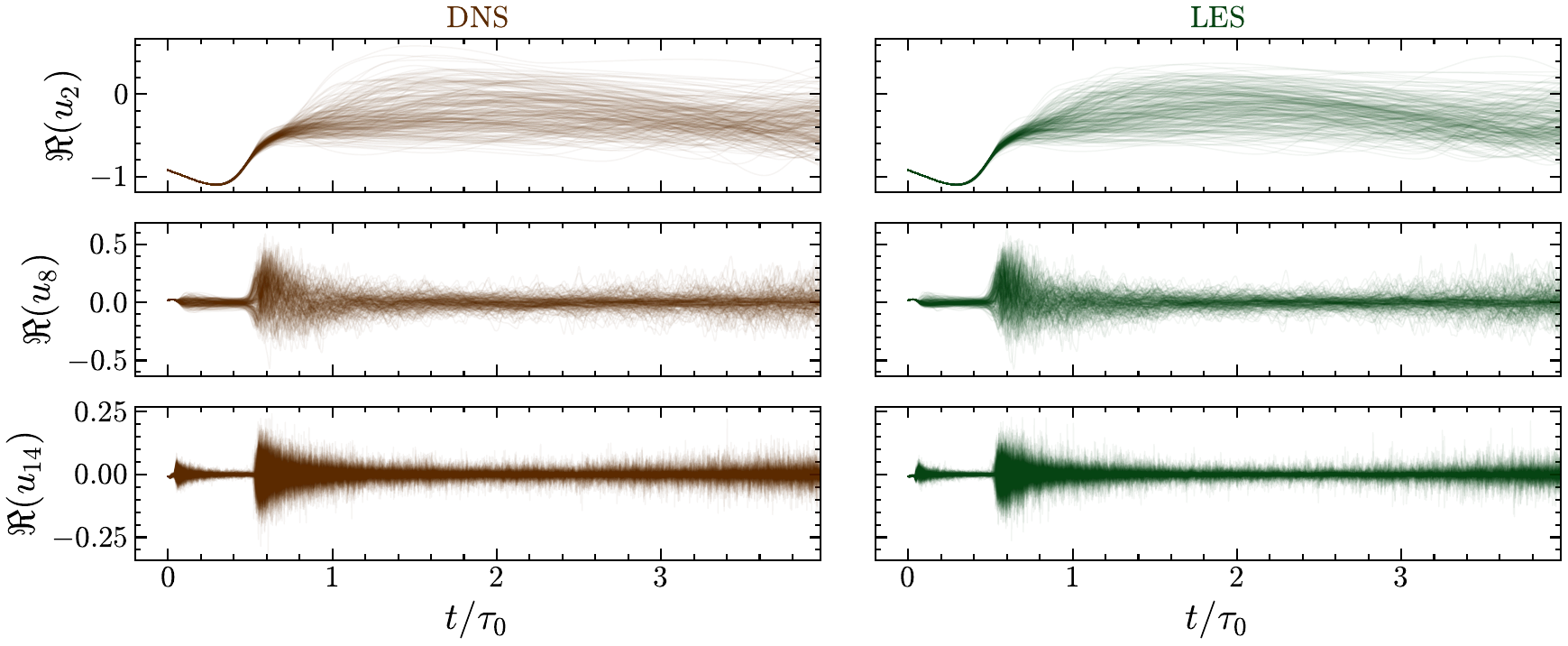}
    \vspace*{-3mm}
    \caption{\textbf{Ensemble evolution from identical initial conditions}. Real part of the shell velocities $\Re(u_n)$ for $n=2,8,14$ versus $t/\tau_0$. DNS (left) uses Landau--Lifshitz fluctuating hydrodynamics with different noise realizations, while LES (right) uses independent realizations of the stochastic subgrid closure. The comparable ensemble spreading across scales illustrates how stochastic closures recover DNS-level uncertainty growth in reduced models.}
    \label{fig:fig2_trajs}
\end{figure*}

We begin with a qualitative comparison of ensemble evolution. Fig.~\ref{fig:fig2_trajs} shows the real part of the shell velocities for shells $n=2,8,14$ for DNS and LES. As expected from the scale-dependent eddy turnover times, ensembles decorrelate rapidly at large $n$ (near the cutoff) while the low-wavenumber shells spread more slowly. Despite the drastic reduction in degrees of freedom, the stochastic LES reproduces a comparable spreading of trajectories across inertial and near-cutoff scales, indicating that sustained stochastic forcing at unresolved scales can recover DNS-level uncertainty propagation.

\begin{figure}[htb]
    \centering
    \includegraphics[width=0.99\linewidth]{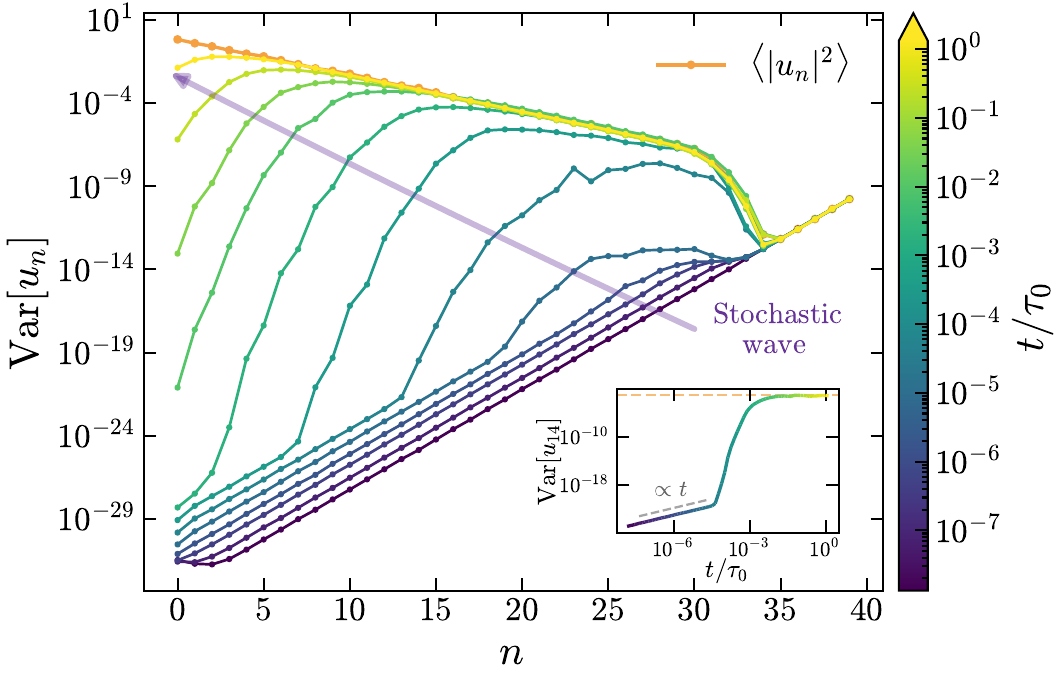}
    \vspace*{-7mm}
\caption{\textbf{Spatiotemporal propagation of uncertainty}. Variance $\mathrm{Var}[u_n]$ versus shell index $n$ at different times for Landau--Lifshitz fluctuating hydrodynamics (color-coded by $t/\tau_0$). At early times, variance is confined to small scales and follows the thermal $k^2$ scaling. A stochastic wave of uncertainty then propagates upscale, realizing the inverse cascade of error anticipated by Lorenz~\cite{lorenz69}, and reaches the largest scales within one turnover time $\tau_0$. The orange curve shows the energy spectrum $\langle|u_n|^2\rangle$, which sets the saturation level of variance at late times. Inset: variance of the cutoff shell, illustrating initial linear growth followed by rapid nonlinear amplification and saturation.}
\label{fig:fig3_var}
\end{figure}

To understand the origin of this behavior, we next examine uncertainty propagation in fully-resolved simulations. Fig.~\ref{fig:fig3_var} shows the spatiotemporal evolution of $\mathrm{Var}[u_n]$ for DNS with fluctuating hydrodynamics (Eq.~\eqref{eq:sabra_ll}). At early times, variance is localised at small scales and follows the thermal equilibrium scaling $\mathrm{Var}[u_n]\propto k_n^2$. As time progresses, uncertainty propagates upscale in the form of a stochastic wave, realizing the inverse cascade of error anticipated by Lorenz~\cite{lorenz69}. Within one large-eddy turnover time $\tau_0$, variance reaches the largest scales and saturates at the level set by the energy spectrum $\langle|u_n|^2\rangle$. The inset illustrates the temporal evolution of the cutoff shell variance, showing the initial linear growth $\mathrm{Var}[u_n]\sim t$ associated with independent noise increments, followed by rapid nonlinear amplification and saturation.

\begin{figure*}[htb]
    \centering
    \includegraphics[width=0.99\linewidth]{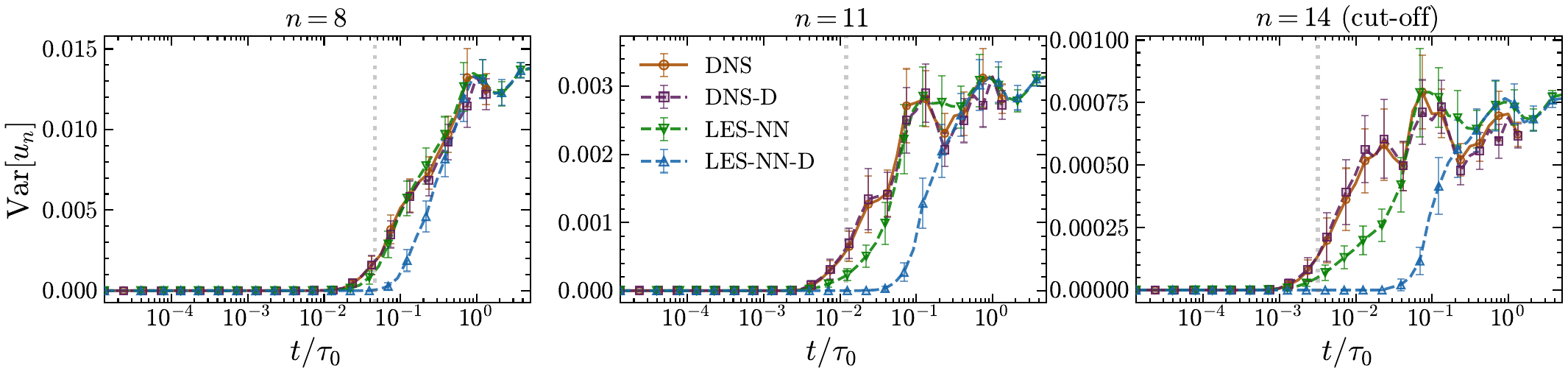}
    \vspace*{-3mm}
\caption{\textbf{Does stochasticity matter?} Time evolution of the variance $\mathrm{Var}[u_n]$ for shells $n=8,11,14$. Shown are Landau--Lifshitz fluctuating hydrodynamics (DNS), the same but with noise applied only at the initial step followed by deterministic evolution (DNS--D), the LES with stochastic closure (LES--NN), and its deterministic counterpart with only an initial step with the stochastic closure (LES--NN--D). In the fully resolved DNS, deterministic chaos suffices to recover the correct variance growth after the initial perturbation. In contrast, for the reduced LES system, deterministic evolution alone leads to a pronounced delay in variance growth. Sustained stochastic forcing in the closure is therefore essential to correctly capture uncertainty propagation at a coarse resolution. The dashed gray vertical lines denote the characteristic time scale of that shell $\tau_n\sim k_n^{-2/3}$.}
    \label{fig:fig4_comp}
\end{figure*}

We then assess whether reduced models reproduce this uncertainty growth quantitatively. Fig.~\ref{fig:fig4_comp} compares the time evolution of $\mathrm{Var}[u_n]$ for DNS and LES at shells $n=8,11,14$. In the fully resolved system, DNS with continuous stochastic forcing and DNS with noise applied only at $t=0$ (DNS--D) quickly collapse onto the same variance evolution once nonlinear transfer dominates. The initial linear-in-time regime associated with independent noise increments occurs at earlier times than those shown here (see the inset of Fig.~\ref{fig:fig3_var}). This agreement indicates that, when all dynamically relevant scales are retained, deterministic chaos rapidly amplifies a single microscopic perturbation and recovers the correct finite-time uncertainty growth.

The LES behaves qualitatively differently. While the stochastic LES--NN closure follows the DNS variance growth reasonably well (with a modest delay), its deterministic counterpart (LES--NN--D), obtained by applying the stochastic closure for one step and then setting $\sigma_n=0$ (without retraining, so that the drift network is identical), exhibits a pronounced delay in variance growth. The delay increases towards the cutoff and persists over times close to $\tau_0$, demonstrating that deterministic evolution alone is insufficient to transport uncertainty across scales in truncated dynamics. The robustness of this conclusion is supported by two additional tests: Appendix~D repeats the analysis with a phenomenological stochastic closure, and Appendix~C explores alternative prescriptions for initial condition perturbations for the deterministic LES.

For completeness, standard deterministic diagnostics of the reduced models (structure functions, flatness, and multiplier PDFs) are reported in Appendix~E.

%-----------------------------------------
%-----------------------------------------
\section{Discussion and outlook}
\label{sec:discussion}

This work addresses a basic question in predictive turbulence modelling: can a reduced model reproduce not only mean statistics but also the \emph{finite-time} growth and propagation of uncertainty across scales? We study this problem in a controlled Sabra shell-model setting by comparing (i) a Landau--Lifshitz fluctuating-hydrodynamics~\cite{landau_lifshitz_fluid_1959} reference (DNS), (ii) a stochastic Langevin-type LES closure, and (iii) deterministic reduced models in which uncertainty is introduced only through initial condition perturbations. The results show a clear dichotomy. In the fully resolved system, a single microscopic perturbation is rapidly amplified and redistributed by deterministic nonlinear dynamics, so that DNS with continuous-in-time noise and deterministic DNS with noise applied only at initialization converge to the same variance growth after a short transient. In contrast, once the dynamics are truncated, deterministic closures exhibit a pronounced delay and suppression of variance growth. This behaviour is consistent with a stochastic ``wave'' of uncertainty that propagates upscale from the dissipation range, realizing the inverse cascade of error anticipated by Lorenz~\cite{lorenz69}: in the fluctuating-hydrodynamics reference, fluctuations seeded in the far-dissipation tail reach large scales within $\mathcal O(\tau_0)$~\cite{bandak_prl_2024}, whereas deterministic reduced models suppress the continual near-cutoff injection required to sustain this transport. The resulting rapid, noise-triggered amplification is closely connected to phenomena of spontaneous stochasticity~\cite{falkovich_particles_2001, thalabard_butterfly_2020}. Recent work has further clarified this connection by deriving, in an exactly renormalisable multiscale lattice model, an explicit link between spontaneous stochasticity and stochastic LES modelling under a renormalisation-group flow~\cite{Peng2025}. In that setting, integrating out early times and small scales yields effective large-scale initial data and subgrid variables that converge to universal, i.i.d. stochastic distributions, independent of the microscopic noise. This provides rigorous support for the interpretation that coarse-graining in multiscale chaotic systems naturally induces robust stochastic reduced dynamics.

Overall, the present work supports a simple conclusion: when turbulence is modelled at reduced resolution, stochasticity in the closure is essential to reproduce physically correct uncertainty growth over $\mathcal O(\tau_0)$ times. More generally, this mechanism is not specific to turbulence but to chaotic, multiscale systems represented by truncated dynamics. Whenever unresolved degrees of freedom remain dynamically active, deterministic reduced models that inject uncertainty only through initial perturbations can become spuriously overconfident. This situation arises most transparently in turbulent transport problems—such as geophysical circulation and weather–climate models~\cite{leith1978predictability, krishnamurthy2019predictability, palmer_stochastic_2019} and cosmological galaxy-formation simulations~\cite{keller_chaos_2019, genel_butterfly_2019}—but similar effects are expected in multiscale reaction–advection–diffusion systems (e.g. combustion~\cite{peters2001turbulent}, atmospheric chemistry~\cite{brasseur2017modeling}, cloud microphysics~\cite{morrison2020confronting}), threshold-driven dynamics (fires, landslides, fault systems~\cite{hergarten2002self, sornette2006critical}), pattern-forming instabilities~\cite{cross1993pattern}, and coarse-grained many-body systems where eliminating fast variables induces effective noise and memory in the sense of Mori–Zwanzig~\cite{MZ_2010, pavliotis2008multiscale}. 

These considerations are directly relevant to predictability studies in climate, where forecast skill is commonly assessed through error-growth experiments and Lyapunov concepts~\cite{leith1978predictability, krishnamurthy2019predictability}. Deterministic LES may overestimate predictability if continual subgrid uncertainty injection is neglected, consistent with the practical success of stochastic parameterisations in ensemble forecasting~\cite{palmer_stochastic_2019}. Stochastic parameterisations have also been argued to permit reliable predictions at lower resolution and reduced arithmetic precision~\cite{palmer_stochastic_2019}, highlighting potential computational advantages as well. The implications extend to ensemble-based data assimilation~\cite{DA_2018}: in ensemble Kalman filters~\cite{Evensen2003}, under-dispersive forecast models can yield collapsed covariances and overconfident analyses, typically corrected by ad hoc inflation or localisation~\cite{gottwald2013mechanism, raanes2019adaptive}. Closely related issues arise in cosmological simulations of galaxy formation, which are chaotic, strongly multiscale, and necessarily under-resolved; even minute perturbations can produce macroscopically distinct outcomes~\cite{keller_chaos_2019, genel_butterfly_2019}, underscoring the importance of representing unresolved processes in a stochastic rather than purely deterministic manner. 

A final outlook direction concerns how stochasticity should be modelled. Our closures adopt a deliberately minimal structure: additive, Markovian forcing with amplitudes fixed from dimensional considerations and statistics of the fully resolved system. This is sufficient to partially restore the correct variance growth, but real subgrid dynamics is neither strictly Gaussian nor memoryless, and its conditional statistics depend on the resolved state. Recent advances in generative modelling~\cite{albergo_2025} offer promising tools to learn such conditional distributions, but most current approaches focus on one-shot sampling; turning them into stable, autoregressive generators that can be embedded efficiently in a numerical solver remains an open and computationally demanding challenge.

%-----------------------------------------
%-----------------------------------------
\acknowledgments
\raggedbottom
This research was supported by European Union’s HORIZON MSCA Doctoral Networks programme under Grant Agreement No. 101072344, project AQTIVATE (Advanced computing, QuanTum algorIthms and data-driVen Approaches for science, Technology and Engineering), the European Research Council (ERC) under the European Union’s Horizon 2020 research and innovation programme Smart-TURB (Grant Agreement No. 882340). A.A.M. was supported by the CAPES MATH-AmSud project CHA2MAN.

\emph{Data availability statement}. All code developed in this work will be made openly available upon publication at \href{https://github.com/andremfreitas/stochastic_closure_sm}{\texttt{github.com/andremfreitas/stochastic\_closure\_sm}}.
%-----------------------------------------
%-----------------------------------------
%\nocite{*}
%\clearpage
%\pagebreak
\bibliographystyle{eplbib}
\bibliography{refs}
\clearpage
\pagebreak
%-----------------------------------------
%-----------------------------------------

%-----------------------------------------
%-----------------------------------------
%-----------------------------------------
%-----------------------------------------
\section{Appendix A: Numerical integration scheme}
\label{sec:appendix_numerical_method}
We integrate the Landau--Lifshitz Sabra model, Eq. ~\eqref{eq:sabra_ll}, using a first-order Lie--Trotter splitting~\cite{hairer2006gni} between the deterministic drift and the (additive) stochastic forcing. Writing the dynamics in It\^o form,
\begin{align}
&\mathrm d u_n
=
\Bigl(\mathcal N_n(u) - \nu k_n^2 u_n\Bigr)\,\mathrm dt
+ \tag{A.1a} 
\sigma_n\,\mathrm dW_n,
\\
&\sigma_n=\sqrt{\Theta}\,k_n^{1+\alpha/2},
\tag{A.1b}
\end{align}
where $\mathcal N_n(u)$ denotes the quadratic Sabra nonlinearity together with the deterministic large-scale forcing, and $\{W_n\}_{n=0}^{N-1}$ are independent complex Wiener processes.

The complex white noises $\xi_n(t)$ in Eq.~\eqref{eq:sabra_ll} are related to $W_n$ by $\xi_n(t)\,\mathrm dt=\mathrm dW_n(t)$ (in distribution), with normalisation
\begin{equation}
\big\langle \mathrm dW_n^\ast(t)\,\mathrm dW_m(t)\big\rangle
=
2\,\delta_{nm}\,\mathrm dt,
\tag{A.2}
\end{equation}
equivalently $\langle \xi_n^\ast(t)\,\xi_m(t')\rangle = 2\,\delta_{nm}\,\delta(t-t')$.

Over one time step $\Delta t$, the Lie--Trotter split-step update is defined by composing the flow of the drift equation
\begin{equation}
\dot u_n = \mathcal N_n(u) - \nu k_n^2 u_n,
\tag{A.3}
\end{equation}
with the flow of the noise-only equation
\begin{equation}
\mathrm d u_n = \sigma_n\,\mathrm dW_n .
\tag{A.4}
\end{equation}
Since the diffusion is additive (independent of $u$), the stochastic subproblem~(A.4) is solved exactly and yields
\begin{align}
&u_n(t+\Delta t)=u_n^{\mathrm{drift}}(t+\Delta t)+\sigma_n\,\Delta W_n,
\tag{A.5a} \\
&\Delta W_n := W_n(t+\Delta t)-W_n(t).
\tag{A.5b}
\end{align}
Thus, the stochastic contribution is an exact additive-noise increment (equivalently, an Euler--Maruyama~\cite{kloeden1992nsde} update applied to the noise-only subproblem).

To compute the drift map $u^{\mathrm{drift}}(t+\Delta t)$, we treat the linear viscous term exactly with an integrating factor~\cite{IFM_2002} and integrate the remaining nonlinear term using a classical fourth-order Runge--Kutta method~\cite{hairer2006gni}. Defining the stages
\begin{align}
K_1 &= \Delta t\,\mathcal N_n\!\left(u(t)\right), \tag{A.6a}\\
K_2 &= \Delta t\,\mathcal N_n\!\left(
\mathrm e^{-\nu k_n^2 \Delta t/2}\bigl(u(t)+\tfrac12 K_1\bigr)
\right), \tag{A.6b}\\
K_3 &= \Delta t\,\mathcal N_n\!\left(
\mathrm e^{-\nu k_n^2 \Delta t/2}\bigl(u(t)+\tfrac12 K_2\bigr)
\right), \tag{A.6c}\\
K_4 &= \Delta t\,\mathcal N_n\!\left(
\mathrm e^{-\nu k_n^2 \Delta t}\bigl(u(t)+K_3\bigr)
\right), \tag{A.6d}
\end{align}
the drift update reads
\begin{equation}
u_n^{\mathrm{drift}}(t+\Delta t)
=
\mathrm e^{-\nu k_n^2 \Delta t}
\Bigl[
u_n(t)+\tfrac16\bigl(K_1+2K_2+2K_3+K_4\bigr)
\Bigr].
\tag{A.7}
\end{equation}
The full split-step method is then obtained by adding the stochastic increment as in~(A.5).

The complex Wiener increments are sampled as
\begin{equation}
\Delta W_n = \sqrt{\Delta t}\,(X_n+iY_n),
\qquad
X_n,Y_n \stackrel{\text{i.i.d.}}{\sim}\mathcal N(0,1),
\tag{A.8}
\end{equation}
which ensures $\langle \Delta W_n^\ast\Delta W_m\rangle = 2\,\delta_{nm}\,\Delta t$ and consistency with~(A.2). The corresponding one-step variance of the stochastic increment is
\begin{equation}
\left\langle \bigl|\sigma_n\,\Delta W_n\bigr|^2 \right\rangle
=
2\,\sigma_n^2\,\Delta t
=
2\,\Theta\,k_n^{2+\alpha}\,\Delta t.
\tag{A.9}
\end{equation}

The time step is fixed to $\Delta t=10^{-8}\approx\tau_\eta/180$, ensuring resolution of the fastest dissipative time scales.

%-----------------------------------------
%-----------------------------------------
\section{Appendix B: Training details}
\label{sec:appendix_training}

We briefly summarize the solver-in-the-loop~\cite{um2020sol} training procedure used to learn the stochastic closure, following and extending the methodology introduced in Refs.~\cite{freitas24,freitas25}. Training is performed by embedding the closure directly into the time integration of the reduced shell model and optimizing the network parameters through multi-step trajectory matching.

Training data are generated from fully resolved deterministic simulations of the Sabra model with $N=40$ shells at Reynolds number $\mathrm{Re}\approx10^{12}$. The reduced model retains the shells up until $s=14$, advanced with time step $\Delta\tilde t=10^{-5}$, while reference trajectories (256 in total) use $\Delta t=10^{-8}$. Training trajectories span $T_{\mathrm{train}}=1.65\,\tau_0$, and testing trajectories span $T_{\mathrm{test}}=3.31\,\tau_0$, where $\tau_0$ is the large-eddy turnover time.

A single training iteration is summarized in Algorithm~\ref{alg:training}, where $u^{<}_t=(u_0,\dots,u_s)$ is the resolved state, $\Phi_{\Delta\tilde t}$ is an RK4 step of the resolved dynamics, $\hat u^{>}_t=\mathrm{NN}_\theta(u^{<}_t)$ provides the subgrid shells used to build the closure $\mathcal C$ acting only on shells $s-1,s$, and $\Sigma\,\xi_t$ is the fixed-amplitude complex Gaussian forcing. Starting from a batch of resolved initial conditions, the neural network predicts the deterministic subgrid drift acting on the last resolved shells. The reduced dynamics are then advanced for $\texttt{m}=100\Delta\tilde t$ in-the-loop steps using the solver-in-the-loop scheme. The resulting trajectories are compared with the reference solution through a multi-step loss, gradients are computed by backpropagation through the solver, and the network parameters are updated using gradient-based optimization.
\begin{algorithm}[htb]
\caption{Training loop (single iteration).}
\label{alg:training}
\begin{algorithmic}[1]
\State Sample batch of initial resolved states $u^{<}_0$
\State Initialize gradient tape
\For{$t=0$ to $m-1$}
    \State Predict subgrid drift terms $\hat u^{>}_t \gets \mathrm{NN}_\theta(u^{<}_t)$
    \State Sample noise $\xi_t \sim \mathcal N_{\mathbb C}(0,1)$
    \State $u^{<}_{t+1} \gets \Phi_{\Delta\tilde t}(u^{<}_t)
    + \Delta\tilde t\,\mathcal C(u^{<}_t,\hat u^{>}_t)
    + \sqrt{\Delta\tilde t}\,\Sigma\,\xi_t$
\EndFor
\State Compute loss $\mathcal L$ against reference trajectories
\State Update parameters $\theta$ using $\nabla_\theta \mathcal L$
\end{algorithmic}
\end{algorithm}
% \begin{table}[htb]
%     \centering
%     \caption{Values of the parameters used in the numerical experiments.}
%     \begin{tabular}{llp{0.4\columnwidth}}
%         \toprule
%         \textbf{Parameter} & \textbf{Value} & \textbf{Description} \\
%         \midrule
%         $\nu$ & $ 10^{-12}$ & Viscosity \\
%         $\mathrm{Re}$ & $\approx 10^{12}$ & Reynolds number \\
%         $\epsilon$ & $0.5$ & Forcing amplitude \\
%         $N$ & $40$ & Number of shells \\
%         $N_\eta$ & $30$ & Kolmogorov scale \\
%         $N_\mathrm{c}$ & $14$ & Subgrid cutoff scale \\
%         $\tau_0$ & $7.553 \times 10^{-1}$ & Eddy turnover time at the integral scale \\
%         $\tau_\eta$ & $1.8367 \times 10^{-6}$ & Eddy turnover time at the dissipative scale \\
%         $\Delta t$ & $1 \times 10^{-8}$ & Timestep of ground truth simulation \\
%         $\Delta \tilde{t}$ & $1 \times 10^{-5}$ & Timestep of LES–NN model \\
%         $N_\mathrm{data}$ & $256$ & Number of initial conditions in the dataset \\
%         $N_\mathrm{batch}$ & $1024$ & Batch size used for training \\
%         $T_\mathrm{train}$ & $1.65\,\tau_0$ & Integration time of training dataset \\
%         $T_\mathrm{test}$ & $3.31\,\tau_0$ & Integration time of test dataset \\
%         $\texttt{msteps}$ & $100$ & Time steps in-the-loop per training iteration \\ 
%         \bottomrule
%     \end{tabular}
%     \label{tab:params}
% \end{table}

%-----------------------------------------
%-----------------------------------------
\section{Appendix C: Initial condition ensembles for deterministic closures}
\label{sec:appendix_ic}

In the main text, we assess uncertainty growth in reduced turbulent models by comparing stochastic closures with deterministic closures evolved from ensembles of (perturbed) initial conditions. This appendix examines alternative prescriptions for constructing such initial ensembles without a stochastic closure and shows that they still fail to reproduce the correct variance growth in the truncated system.

For the fully resolved dynamics, deterministic chaos is ultimately sufficient. After an initial trivial transient, during which $\mathrm{Var}[u_n(t)]\propto t$ due to additive noise increments
$u_n(t+\Delta t)=u_n(t)+\sqrt{\Delta t}\,\sigma_n\xi_n$
(with $\mathbb E|u_n(t)|^2\sim\sigma_n^2 t$ for independent increments), the variance growth obtained from continuously forced Landau--Lifshitz dynamics collapses onto that obtained from a single noisy initialization followed by deterministic evolution. This collapse occurs once nonlinear transfer dominates the early linear-in-time accumulation. The equivalence breaks down after truncation. Meaning that in the truncated case estimating uncertainty is crucial, even though unclear how to do so. For this, we  here consider three deterministic initial ensemble heuristics for the reduced LES model (all followed by purely deterministic evolution with the same learned drift term).  
(i) \emph{Kolmogororov-based perturbations} (LES--NN--D--$\eta$): perturb the last resolved shells as
$u_n\mapsto u_n+\varepsilon_\eta U_\eta \zeta_n$ for $n$ near the cutoff, where $\zeta_n$ are i.i.d.\ complex Gaussians of unit variance and $U_\eta$ is estimated by inertial-range scaling,
$U_\eta \sim U_c (k_\eta/k_c)^{-1/3}$, with $U_c$ measured at the cutoff shell.  
(ii) \emph{Round-off perturbations} (LES--NN--D--P): apply multiplicative machine-precision noise
$u_n\mapsto u_n(1+\varepsilon_{\mathrm{mach}}\zeta_n)$ with $\varepsilon_{\mathrm{mach}}\sim10^{-16}$, representing the minimal uncertainty from finite (double) precision.  
(iii) \emph{Single-step with stochastic closure} (LES--NN--D--S): apply the stochastic closure one step and then switch to a purely deterministic closure for the rest of the evolution.

\begin{figure*}[htb]
    \centering
    \includegraphics[width=1.0\linewidth]{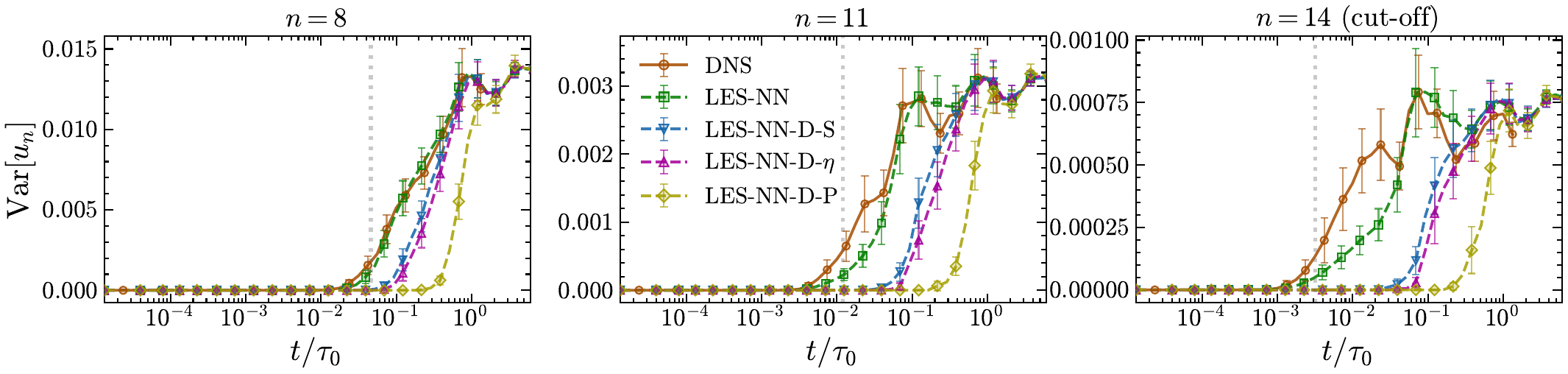}
    \caption{\textbf{Different initial condition perturbations for the LES evolution with a deterministic closure}. 
    Ensemble variance $\mathrm{Var}[u_n]$ versus $t/\tau_0$ for shells $n=8,11,14$, comparing the fluctuating hydrodynamics reference (DNS), the stochastic reduced model (LES--NN), and deterministic reduced model evolution obtained by different prescriptions for the initial condition perturbations: one step with the stochastic closure followed by deterministic evolution (LES--NN--D--S), Kolmogorov-based perturbations applied near the cutoff (LES--NN--D--$\eta$), and machine-precision round-off perturbations (LES--NN--D--P). The dashed gray vertical lines denote the characteristic time scale of that shell $\tau_n\sim k_n^{-2/3}$.}
    \label{fig:diff_perturb}
\end{figure*}

Fig.~\ref{fig:diff_perturb} compares variance growth in time for shells $n=8,11,14$ between fluctuating hydrodynamics, the stochastic LES and the deterministic LES with different initial condition perturbations. Although all deterministic ensembles eventually decorrelate and match the correct asymptotic variance, their variance growth is systematically delayed. As expected, injecting uncertainty solely through initial conditions does not propagate efficiently across scales in the truncated system, leading to an artificial suppression of variance on times of order the large-eddy turnover time.

These results reinforce the main conclusion: unlike in fully resolved simulations, deterministic chaos seeded by initial perturbations is not sufficient to model uncertainty growth in reduced systems, and an explicit stochastic forcing in the closure is essential to represent the amplification and transport of uncertainty across scales.

\begin{figure*}[h]
    \centering
    \includegraphics[width=1.0\linewidth]{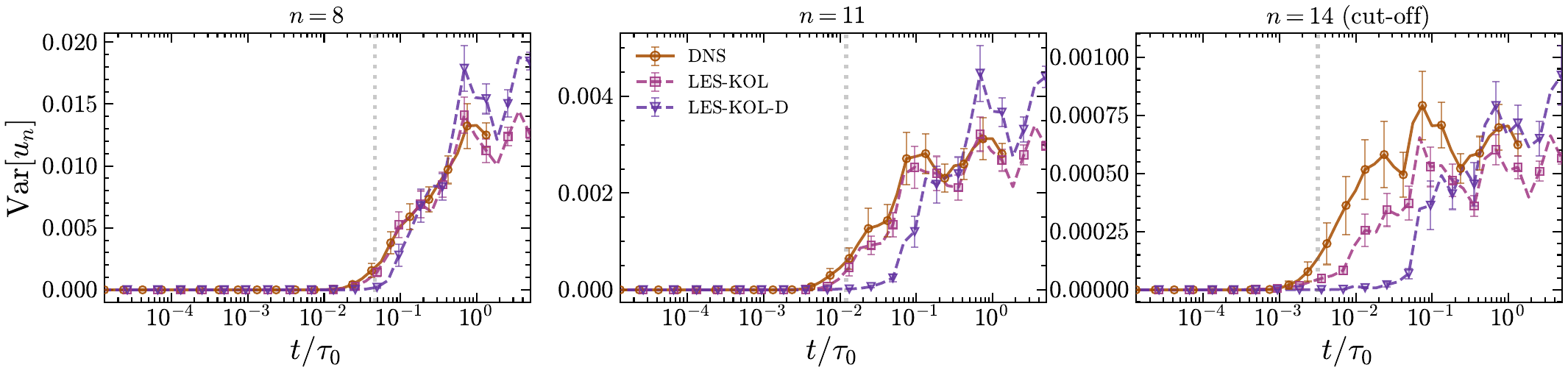}
    \caption{\textbf{Phenomenological stochastic closure: variance growth.} Ensemble variance $\mathrm{Var}[u_n]$ versus $t/\tau_0$ for shells $n=8,11,14$ comparing the fluctuating hydrodynamics reference (DNS), a stochastic phenomenological closure based on stochastic multipliers (LES--KOL), and its deterministic counterpart (LES--KOL--D) obtained by sampling the multipliers once at $t=t_0$ from the stochastic closure and then for the subsequent steps freezing them at $\lambda^{-1/3}$ and $\pi/2$ for the amplitude and phase multipliers, respectively. The dashed gray vertical lines denote the characteristic time scale of that shell $\tau_n\sim k_n^{-2/3}$.}
    \label{fig:smk}
\end{figure*}

%-----------------------------------------
%-----------------------------------------
\section{Appendix D: Phenomenological closure}
\label{sec:appendix_smk}
%-----------------------------------------

To assess the robustness of our conclusions beyond neural networ based closures, we consider a stochastic extension of the phenomenological subgrid model introduced in Ref.~\cite{biferale_AAM_Parisi}.

For the Sabra shell model, the closure is formulated in terms of amplitude and phase multipliers~\cite{Kolmogorov_1962}. These are defined as
\begin{align}
w_n &= \left| \frac{u_n}{u_{n-1}} \right| , \tag{D.1}\label{eq:amp_mult}\\
\Delta_n &= \arg(u_n) - \arg(u_{n-1}) - \arg(u_{n-2}). \tag{D.2}\label{eq:phase_mult}
\end{align}
In the inertial range, the amplitude multipliers fluctuate around the Kolmogorov scaling $w_n\sim\lambda^{-1/3}$~\cite{kolmogorov1941local}, while the phase multipliers have a distribution peaked at $\pi/2$, corresponding to a forward energy cascade~\cite{biferale_AAM_Parisi}.

In the deterministic closure, unresolved shells are reconstructed by fixing these multipliers to their most probable values. In contrast, here we introduce fluctuations around these fixed points. For the first two unresolved shells $s+1$ and $s+2$, we write
\begin{align}
\Delta_{s+1}(t) &= \frac{\pi}{2} + x_1(t), \qquad
\Delta_{s+2}(t) = \frac{\pi}{2} + x_2(t), \tag{D.3a}\\
w_{s+1}(t) &= \lambda^{-1/3} \, \mathrm e^{\eta_1(t)}, \qquad
w_{s+2}(t) = \lambda^{-1/3} \, \mathrm e^{\eta_2(t)}. \tag{D.3b}
\end{align}
Here $x_i(t)$ represent stochastic phase fluctuations around the peak $\pi/2$, while $\eta_i(t)$ describe lognormal amplitude fluctuations around Kolmogorov scaling. The amplitudes are constructed such that
\begin{equation}
\mathbb E\!\left[\mathrm e^{\eta_i}\right] = 1,
\tag{D.4}
\label{eq:mean_lognormal}
\end{equation}
ensuring that the mean energy flux is unchanged.

Both $x_i$ and $\eta_i$ evolve as independent Ornstein–Uhlenbeck processes,
\begin{align}
\mathrm d x_i &= -\tau_i^{-1} x_i\,\mathrm dt + \sigma_x\,\tau_i^{-1/2}\,\mathrm dW_i^{(x)}, \tag{D.5}\\
\mathrm d \eta_i &= -\tau_i^{-1}(\eta_i-\mu_\eta)\,\mathrm dt
+ \sigma_\eta\,\tau_i^{-1/2}\,\mathrm dW_i^{(\eta)}, \tag{D.6}
\end{align}
with $\mu_\eta=-\sigma_\eta^2/2$ (to satisfy Eq.~\eqref{eq:mean_lognormal}). The correlation times $\tau_i$ are taken from the eddy turnover times at the corresponding shells $\tau_n\sim k_n^{-2/3}$. In this formulation, the only free parameters of the closure are the standard deviations $\sigma_x$ and $\sigma_\eta$, controlling phase and amplitude fluctuations, respectively. We fix $\sigma_x = \sigma_\eta = 0.5$.

At each integration stage, the unresolved shell variables $u_{s+1}$ and $u_{s+2}$ are reconstructed from $(w_i,\Delta_i)$ and injected into the nonlinear Sabra interactions. The resolved shells are then advanced in time using the numerical solver described in Appendix A, yielding a stochastic LES model without any learned components.

Fig.\ref{fig:smk} compares ensemble variance growth for three cases: Landau--Lifshitz fluctuating hydrodynamics (DNS), the stochastic phenomenological closure (LES--KOL), and its deterministic counterpart (LES--KOL--D) in which the stochastic multipliers are sampled once and then held fixed at $\lambda^{-1/3}$ and $\pi/2$. Consistently with the neural network closure, LES--KOL closely reproduces the timing and magnitude of DNS variance growth across shells with a very slight delay, whereas LES--KOL--D exhibits a strong systematic delay, most pronounced near the cutoff. This confirms that the delayed uncertainty growth is not an artifact of the learning architecture but a generic consequence of removing sustained stochastic subgrid excitation in truncated dynamics. We note that introducing multiplier fluctuations can also modify the mean effectiveness of the closure (by relaxing the strictly fixed Kolmogorov scaling and phase multiplier values), but the predictability signature remains clearly visible and in line with what was observed before.

%-----------------------------------------
%-----------------------------------------
\section{Appendix E: Deterministic observables}
\label{sec:appendix_det_obs}
%-----------------------------------------
For completeness, we assess standard deterministic statistics produced by the reduced models. We compare the fully resolved simulation (DNS) with the LES with the data-driven Langevin closure, Eq.~\eqref{eq:subgrid_langevin}, and with the stochastic phenomenological (multiplier-based) closure described in Appendix~D.

\begin{figure}[htb]
    \centering
    \includegraphics[width=0.8\linewidth]{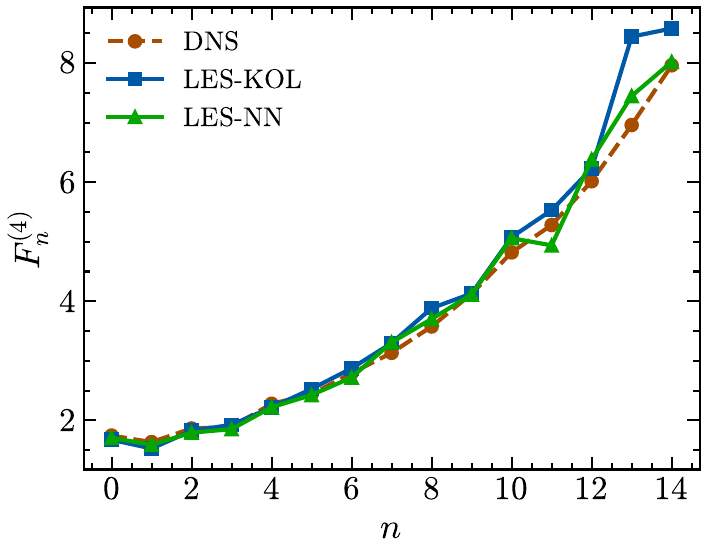}
    \caption{\textbf{Intermittency in the reduced models.} Flatness $F_n^{(4)}$ versus shell index $n$, comparing DNS with LES using the phenomenological stochastic closure (LES--KOL) and the data-driven Langevin closure (LES--NN).}
    \label{fig:flatness}
\end{figure}

\begin{figure}[htb]
    \centering
    \includegraphics[width=0.8\linewidth]{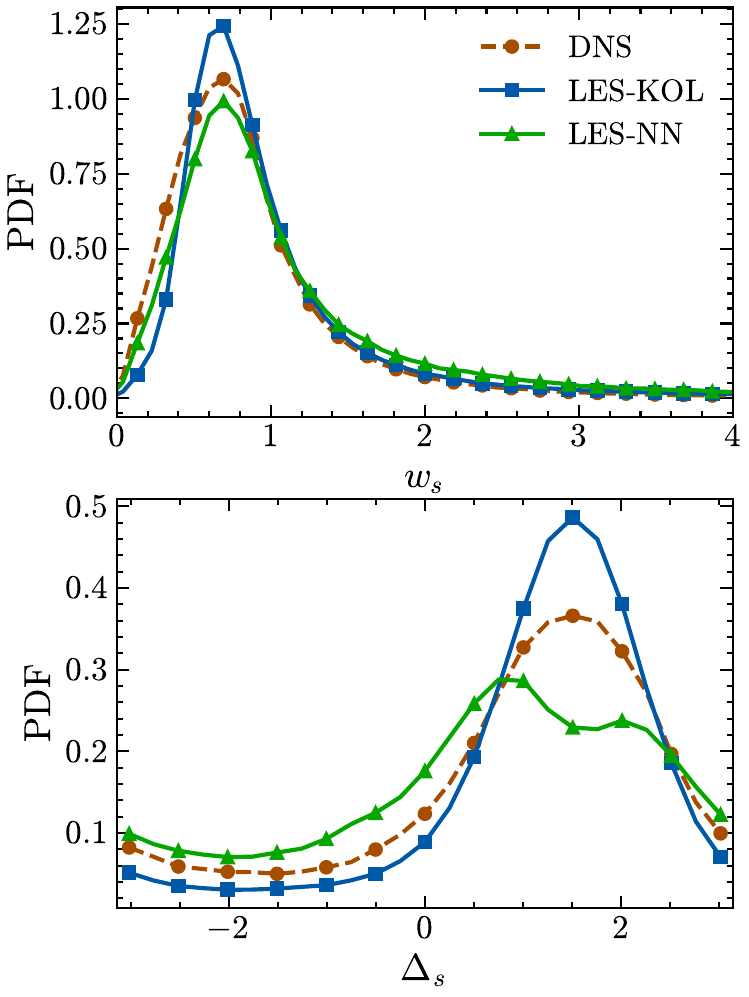}
    \caption{\textbf{Multiplier statistics at the cutoff.} Probability density functions of the amplitude $w_s$ and phase $\Delta_s$ multipliers at the cutoff shell $s$, comparing DNS with LES using the phenomenological stochastic closure (LES--KOL) and the data-driven Langevin closure (LES--NN).}
    \label{fig:multipliers_comp}
\end{figure}

We consider the $p$th-order structure functions
\begin{equation}
S_n^{(p)}=\bigl\langle |u_n|^p \bigr\rangle,
\tag{E.1}
\end{equation}
and the corresponding flatness
\begin{equation}
F_n^{(p)}=\frac{S_n^{(p)}}{\bigl(S_n^{(2)}\bigr)^{p/2}}.
\tag{E.2}
\end{equation}
Fig.~\ref{fig:flatness} shows $F_n^{(4)}$ across shells, providing a quantitative measure of intermittency. We also report the probability density functions of the amplitude and phase multipliers, $w_s$ and $\Delta_s$, defined in Eqs.~\eqref{eq:amp_mult}--\eqref{eq:phase_mult}, evaluated at the cutoff shell $s$ (Fig.~\ref{fig:multipliers_comp}). For a more detailed analysis of such observables for data-driven closures, see Refs.~\cite{freitas24,freitas25}; for the phenomenological multiplier closure, see Ref.~\cite{biferale_AAM_Parisi}.

\raggedbottom

\end{document}